%% file: beta.tex
\documentclass[12pt,a4paper]{article}  
\parskip=\medskipamount		
\usepackage{amsfonts,amsmath,amssymb}
\usepackage{bbm} 
\usepackage{arydshln}   
\usepackage[mathscr]{euscript} 
\usepackage{subfigure}
\usepackage[square, comma, numbers, sort&compress]{natbib}
\usepackage[pdftex]{graphicx,color}
\usepackage[pdftex,bookmarks]{hyperref}
\usepackage{hypernat} 
\hypersetup{
		colorlinks=true,pdfstartview=FitH
        bookmarksopen,bookmarksnumbered,citecolor=blue,pdfstartview=FitH,
        pdfauthor = {Simon J. Tyler},
        pdftitle = {Two loop Kahler potential in beta-deformed N=4 SYM theory},
        pdfsubject = {Physics},
        pdfkeywords = {Kahler, supersymmetry, two-loop, beta-deformed},
        pdfcreator = {LaTeX with hyperref package},
        pdfproducer = {pdflatex}
}
\include{preamble}
\numberwithin{equation}{section}
\numberwithin{figure}{section}
\newcommand{\qb}{{\bar q}}
\newcommand{\qm}{{q^{-1}}}
\newcommand{\qbm}{{\bar q^{-1}}}
\newcommand{\hb}{{\bar h}}
\newcommand{\fb}{{\bar\phi}}
\newcommand{\cMg}{\cM_{(g,1)}}
\newcommand{\cMh}{\cM_{(h,q)}}
\newcommand{\Tg}{T_{(g,1)}}
\newcommand{\Th}{T_{(h,q)}}
\newcommand{\Ghl}{{\stackrel{\hookleftarrow}{G}}}
\newcommand{\Ghr}{{\stackrel{\hookrightarrow}{G}}}
\newcommand{\Gb}{{\scG}}
\newcommand{\Gbhl}{{\stackrel{\hookleftarrow}{\scG}}{}}
\newcommand{\Gbhr}{{\stackrel{\hookrightarrow}{\scG}}{}}
\newcommand{\hp}{{\bm\h'\!}}
\newcommand{\vpp}{{\varpi'\!}}
\newcommand{\Ls}{{\rm Ls}}
\newcommand{\Lc}{{\rm Lc}}
\newcommand{\Cl}{{\rm Cl}}
\def\Mp{\big|(q-\qm)(\f_1-\frac\rmi{\sqrt3}\f_2)\big|^2}
\def\Mm{\big|(q-\qm)(\f_1+\frac\rmi{\sqrt3}\f_2)\big|^2}
\def\Ma{|(q-\qm)\f_1+(q+\qm)\f_2|^2}
\def\Mb{|(q+2\qm)\f_1+q\f_2|^2}
\def\Mc{|(q+2\qm)\f_1-q\f_2|^2}
\def\Mat{|(q-\qm)\f_1-(q+\qm)\f_2|^2}
\def\Mbt{|(2q+\qm)\f_1+\qm\f_2|^2}
\def\Mct{|(2q+\qm)\f_1-\qm\f_2|^2}
\def\MaZ{f_q|2\f_2|^2} 
\def\MbZ{f_q|3\f_1+\f_2|^2} 
\def\McZ{f_q|3\f_1-\f_2|^2}
\def\Maz{|2\f_2|^2} 
\def\Mbz{|3\f_1+\f_2|^2} 
 
\begin{document}
\begin{titlepage}

\vspace{30mm}

\begin{center}
{\Large \bf  Two loop K\"ahler potential in 
$\bm \b$-deformed $\bm {\cN = 4}$ SYM theory
}
\end{center}

\begin{center}

{\large  
Simon J. Tyler\footnote{{styler@physics.uwa.edu.au}}
} \\
\vspace{5mm}

\footnotesize{
{\it School of Physics M013, The University of Western Australia\\
35 Stirling Highway, Crawley W.A. 6009, Australia}}  
~\\
\vspace{2mm}
\end{center}

\vspace{5mm}

\pdfbookmark[1]{Abstract}{abstract_bookmark}
\begin{abstract}
\baselineskip=14pt
In $\cN\!=\!2$ superconformal field theories the K\"ahler potential
is known to be tree level exact.  
The $\b$-deformation of $\cN\!=\!4$ $SU(N)$ SYM 
reduces the amount of supersymmetry to $\cN\!=\!1$, 
allowing for non-trivial, superconformal loop corrections 
to the K\"ahler potential.
We analyse the two-loop corrections on the 
Coulomb branch for a complex deformation.
For an arbitrary chiral field in the Cartan subalgebra
we reduce the problem of computing the two-loop K\"ahler potential to
that of diagonalising the mass matrix,
we then present the result in a manifestly superconformal form. 
The mass matrix diagonalisation is performed 
for the case of the chiral background that 
induces the breaking pattern $SU(N)\to SU(N-2) \times U(1)^2$.
Then, for the gauge group $SU(3)$, the K\"ahler potential is explicitly 
computed to the two-loop order.
\end{abstract}
\vfill
\end{titlepage}
\section{Introduction}\label{intro.sect}
The marginal deformations \citep{Leigh:1995ep} of $\cN\!=\!4$ 
supersymmetric Yang-Mills theory (SYM) are a class of $\cN\!=\!1$ superconformal
field theories which have enjoyed a lot of attention
in recent years.  In particular, the $\b$-deformation
has been the subject of intense investigations, 
since its supergravity dual was found
in \citep{Lunin:2005jy}.  Many aspects of the $\b$-deformed theory have
been studied at both the perturbative and nonperturbative level.
In this paper we concentrate only on perturbative aspects.

An important observation of \citep{Leigh:1995ep} is that the 
renormalisation group beta function vanishes 
(the deformation becomes exactly marginal) 
subject to a single, loop corrected, constraint on the deformed couplings.  
The nature of this constraint has been examined in both the perturbative
and nonperturbative windows
using a range of methods and in a variety of limits, e.g. 
\citep{Freedman:2005cg,Penati:2005hp,Rossi:2005mr,Rossi:2006mu,Mauri:2005pa,
Khoze:2005nd,Georgiou:2006np,Ananth:2006ac,Chu:2007pb,Oz:2007qr}
and is still a topic of ongoing discussion
\citep{Elmetti:2006gr,Elmetti:2007up,Elmetti:2007ey,Bork:2007bj,Kazakov:2007dy}.
Despite this wealth of knowledge about the requirements for conformal invariance
in $\b$-deformed theories, the exact functional nature of the 
quantum corrections has received less attention 
\citep{Kuzenko:2005gy,Kuzenko:2007tf,Dorey:2004xm}.
The purpose of this paper is to continue in the vein of
\citep{Kuzenko:2005gy,Kuzenko:2007tf} and investigate the structure of
the two-loop K\"ahler potential in the $\b$-deformed theory.

The K\"ahler potential is a supersymmetric generalisation 
of the effective potential \citep{Coleman:1973jx} and thus it can be used to 
examine the renormalisation effects and vacuum structure of a quantised theory.
Superfield calculations of the one-loop K\"ahler potential in $\cN\!=\!1$ 
superspace are presented 
in \citep{Buchbinder:1993ud,Pickering:1996gt,Grisaru:1996ve},
while two-loop corrections to the Wess-Zumino model 
have been found using superfield methods in, e.g., \citep{Buchbinder:1996cy}.
A computation of the two-loop K\"ahler potential of a
general, non-renormalisable $\cN\!=\!1$ theory was presented in 
\citep{Nibbelink:2005wc}, we will compare with this result 
and discuss its limitations in the conclusion.
In $\cN\!=\!1$ theories the K\"ahler potential is a particularly interesting
sector of the low energy effective action in that it is not constrained 
by holomorphy in the way that the superpotential and gauge potential are. 
This is not the case for $\cN\!=\!2$ theories where the 
non-renormalisation theorems \citep{Seiberg:1993vc, Buchbinder:1997ib} 
imply that the K\"ahler potential receives only 
one-loop corrections and even those vanish in the case of a 
conformally invariant theory \citep{Buchbinder:1997ib}.
This means that the K\"ahler potential of $\b$-deformed $\cN\!=\!4$ 
SYM is purely a product of the deformation.
It is for this reason that we find the K\"ahler potential a particularly
interesting object to examine in the $\b$-deformed SYM theory.

A major technical ingredient of any two-loop effective potential calculation 
is the functional form of the vacuum sunset integral.
In this paper the knowledge of its structure 
is necessary for the presentation of the explicit 
conformal invariance of our results.
The integral has been discussed many times in the literature, e.g.
\citep{Bij:1983bw,Hoogeveen:1985tf,McDonald:2003zj,Ford:1991hw,Ford:1992pn,
Davydychev:1992mt, Broadhurst:1993mw, Ghinculov:1994sd, Davydychev:1995mq, 
Caffo:1998du, Davydychev:1999mq, Martin:2002iu,Nibbelink:2005wc, Kuzenko:2007tf} 
and references therein, 
but to make our discussion both clearer and self-contained, 
we present a new, and we hope simpler, form for the two-loop integral.  
Like the results implied in \citep{Bij:1983bw,Hoogeveen:1985tf,McDonald:2003zj},
the functional form that we find is explicitly symmetric in all three masses 
and holds for all values of the masses, yet our result is much more compact.  
Our derivation is based on the method of characteristics, an approach first 
used in \citep{Ford:1992pn}.

The structure of this paper is as follows:
In section \ref{quant.sect} we review the aspects of the background field
quantisation of $\b$-deformed $\cN\!=\!4$ SYM that are necessary for our 
calculation, including the structure of the mass matrix for an arbitrary
background in the Cartan subalgebra.  Sections \ref{1loop.sect} and
\ref{2loop.sect} are devoted to the calculation of the one and two-loop
K\"ahler potentials respectively.  In section \ref{simpBG.sect}
we make the results explicit by choosing a $SU(3)$-like background.
Finally the appendix contains a review of the structure of the two-loop, 
vacuum sunset diagram.
\section{Quantisation of \texorpdfstring{$\b$}{beta}-deformed 
\texorpdfstring{$\cN\!=\!4$}{N=4} SYM theory}\label{quant.sect}
To keep the following discussion as concise as possible, we only review
the parts of the quantisation process that are necessary for the calculation of 
the K\"ahler potential, see \citep{Kuzenko:2005gy,Kuzenko:2007tf} for 
more details.

Using the superspace conventions of \citep{Buchbinder:1998qv}, 
the classical action for $\b$-deformed $\cN\!=\!4$ $SU(N)$ SYM theory is
\begin{equation*}
        S=\intz \trF\F^\dag_i\F_i + \frac1{g^2}\intc\trF W^2
        +\left\{h\intc\trF({q}\F_1\F_2\F_3-\qm\F_1\F_3\F_2) + \cc \right\}~,
\end{equation*}
where $q=\rme^{\rmi\p\b}$ is the deformation parameter, $g$ is the gauge 
coupling and $h$ is the chiral vertex coupling.  
The undeformed theory corresponds to the limits $q\to1$ and $h\to g$. 
All of the fields transform in the adjoint representation of the gauge group 
and all fields are covariantly (anti)chiral,
\begin{equation} 
 \cDb_\da \cW_\b=0~,\quad  
 \cDb_\da \F_i=0~,\quad i=1,2,3~, 
\end{equation}
where $\cW_\a=\cW_\a^a\,T_a$ and $\F_i=\F_i^a\,T_a$ are 
Lie-algebra valued superfields. The generators $T_a$ 
correspond to the fundamental representation of $SU(N)$. 
Under the condition that the theory remains conformally invariant 
upon quantisation we can write $g$ as a function of $h$ and $q$, 
\citep{Leigh:1995ep}. 
In the general case, this condition is only known to the first few orders in 
the loop expansion. 
For real $\b$ the condition for conformal invariance in the planar (large $N$)
limit is known to be $|h|=g$ to all loops,
\citep{Mauri:2005pa,Ananth:2006ac}.

As noted in \citep{Kuzenko:2007tf}, it is useful to view the above $\cN\!=\!1$
action as a pure $\cN\!=\!2$ SYM theory (described by $\F_1$ and $\cW_\a$) coupled 
to a \emph{deformed hypermultiplet} in the adjoint (described by $\F_{2,3}$).
For if we quantise using only a $\cN\!=\!2$ SYM background lying in the Cartan
subalgebra then we are automatically on the Coulomb branch of the theory 
and all aspects of the deformation are captured in the 
hypermultiplet propagators and chiral cubic vertices.

Using the $\cN\!=\!1$ background field formalism 
\citep{Gates:1983nr} 
we split the dynamical variables
\begin{equation}\label{qtm.bgnd.split}
\F_i\to\F_i+\vf_i~,\quad 
\cD_\a\to\rme^{-gv}\cD_\a\rme^{gv}~,\quad \cDb_\da\to \cDb_\da~,
\end{equation}
with lower-case letters denoting quantum superfields.
To compute the quantum corrections to the 
K\"ahler potential on the Coulomb branch 
we choose the only non-zero background superfield to be $\F_1=\F$ and 
take it to be in the Cartan subalgebra.
We also need to systematically ignore all derivatives that hit the background field.  
Note that the absence of a background gauge field means that the
covariant derivatives reduce to flat ones, i.e. $\cD_A=D_A$.
In general, a background of this type will break the gauge symmetry of the
theory as $SU(N)\to U(1)^{N-1}$.

Gauge fixing with the supersymmetric 't Hooft gauge 
\citep{Ovrut:1981wa,Marcus:1983wb,Binetruy:1983za,Banin:2002mf,Banin:2003es}, 
the quadratic parts of the action are \citep{Kuzenko:2005gy}
\begin{subequations}\label{quad.class.act}\begin{align}
S_{\rm YM}^{(2)}&=-\half\intz\trF\left(v(\Box - |\cMg|^2)v\right)
  +\intz\trF\left(\vf_1^\dag\Box^{-1}(\Box-|\cMg|^2)\vf_1\right)\\
S_{\rm hyp}^{(2)}&=\intz\trF\left(\vf_2^\dag\vf_2+\vf_3^\dag\vf_3\right)
  +\intc\trF\vf_3\cMh\vf_2+\intac\trF\vf_2^\dag \cMh^\dag\vf_3^\dag\\
S_{\rm gh}^{(2)}&=\intz\trF\left(c^\dag\Box^{-1}(\Box-|\cMg|^2)\ct
  -\ct^\dag\Box^{-1}(\Box-|\cMg|^2)c\right)~,
\end{align}\end{subequations}
where we've used the mass operators introduced in \citep{Kuzenko:2005gy}, which 
are elegantly defined by their action on a Lie-algebra valued superfield:
\begin{equation}\label{mass.op.defn}\begin{split}
	\cMh\S&=h({q}\F\S-\qm\S\F)-h\frac{{q}-\qm}{N}\trF(\F\S)\ds1\\
	\cMh^\dag\S&=\hb(\qb\F^\dag\S-\qbm\S\F^\dag)
		-\hb\frac{\qb-\qbm}{N}\trF(\F^\dag\S)\ds1~.
\end{split}\end{equation}

The relevant interactions for the two-loop diagrams of interest are the
cubic couplings%
\begin{subequations}\begin{align} S_{\I}^{(3)}&=
h\intc\trF\left(q\vf_1\vf_2\vf_3-\qm\vf_1\vf_3\vf_2\right)+\cc
 =-\intc(\Th^a)^{bc}\vf_1^a\vf_2^b\vf_3^c-\cc\\
S_{\II}^{(3)}&=
g\intz\trF\left(\vf_i^\dag[v,\vf_i]\right)
=-\intz(\Tg^a)^{bc}\vfb^a_iv^b\vf^c_i~,
\end{align}\end{subequations}
where, following \citep{Kuzenko:2007tf}, we introduce the deformed adjoint
generators
\begin{equation}
(\Th^a)^{bc}=-h\trF(qT^aT^bT^c-\qm T^aT^cT^b)~,
\end{equation}
which enjoy the algebraic properties
\begin{equation}
(\Th^a)^\T=T^a_{(h,-\qm)}~,\quad (\Th^a)^\dag=T^a_{(\hb,\qb)} ~.
\end{equation}
Note that the deformed generators can also be used to give the mass operator
the compact representation $(\cMh)^{ab}=\F^c(T_{(h,\qm)}^c)^{ab}$.

The propagators for the action \eqn{quad.class.act} that are used in
the two-loop calculation below are 
\begin{align}
\rmi \big<v(z)v^\T(z')\big>&=-G_{(g,1)}(z,z') &
\rmi \big<\vf_1(z)\vf_1^\dag(z')\big>&=\frac{\Db^2D^2}{16}G_{(g,1)}(z,z') \\
\rmi \big<\vf_2(z)\vf_2^\dag(z')\big>&=\frac{\Db^2D^2}{16}\Ghl_{(h,q)}(z,z') &
\rmi \big<\vfb_3(z)\vf_3^\T(z')\big>&=\frac{D^2\Db^2}{16}\Ghr_{(h,q)}(z,z') ~,
\end{align}
where all of the fields are treated as adjoint column-vectors, 
in contrast to the Lie-algebraic notation used in defining the action.
The Green's functions are defined by
\begin{align}
\left(\Box-\cMh^\dag\cMh\right)\Ghr_{(h,q)}(z,z')&=-\d^8(z,z')\\
\left(\Box-\cMh\cMh^\dag\right)\Ghl_{(h,q)}(z,z')&=-\d^8(z,z')~,
\end{align}
with the usual, causal boundary conditions.  
As we only have flat derivatives, the above equations are 
most simply solved by moving to momentum space.
In the limit of vanishing deformation the mass matrices commute 
so that the left and right Green's functions coincide: 
$\Ghr_{(g,1)}=\Ghr_{(g,1)}=G_{(g,1)}$.

Throughout this paper we will use dimensional reduction
\citep{Siegel:1979wq} and since we only go to two loops
we do not worry about any possible inconsistencies
\citep{Siegel:1980qs,Avdeev:1981vf}.
This is merely a convenience, as none of the results in this paper
rely on the choice of regularisation scheme and can in fact be
argued at the level of the integrands.

\subsection{Cartan-Weyl basis and the mass operator}\label{CWbMO.sect}
The properties of the mass matrices defined in \eqn{mass.op.defn} 
play a central role in our computations. 
For explicit calculations a convenient choice of basis for our gauge group is
the Cartan-Weyl basis, see e.g. \citep{Barut:1986}.
In this subsection we introduce some notation and a few results that will
be used subsequently.

Any element in \su{N} can be expanded in the Cartan-Weyl basis,
\begin{equation} 
\j = \j^aT_a = \j^{ij}E_{ij}+\j^IH_I~,\quad i\neq j~,
\end{equation}
where $T_a$ is the arbitrary basis used above 
and we choose our Cartan-Weyl basis as the set
\begin{equation}
E_{ij}~,  \quad i\neq j=1,\ldots, N~, 
\qquad H_I ~, \quad I=1,\ldots, N-1~.
\end{equation}
The Cartan-Weyl basis satisfies\footnote{Due to our choice of 
normalisation the Cartan metric is just the Kronecker delta, thus
we can raise and lower the group indices with impunity.} 
\begin{equation}
\tr E_{ij}E_{kl}=\d_{il}\d_{jk}~,
\quad \tr H_I H_J = \d_{IJ}\quad\mbox{and}
\quad \tr E_{ij}H_K=0
\end{equation}
and its elements, defined as matrices in the fundamental representation, are%
\begin{equation} 
(E_{ij})_{kl}=\d_{ik}\d_{jl}~, \qquad 
H_I=\frac1{\sqrt{I(I+1)}}\sum_{i=1}^{I+1}\left(1-i\d_{i(I+1)}\right)E_{ii}~. 
\end{equation}

Since the background is chosen to be in the Cartan subalgebra, 
\[ \F=\f^IH_I:=\F^iE_{ii} ~,\]
the mass matrix is block diagonal when written in the Cartan-Weyl basis
\begin{equation}
\cMh=\bem \cMh^{ijkl} & 0\\0&\cMh^{IJ} \eem
=\bem m^{ki}\d_{il}\d_{jk} & 0\\0&\cMh^{IJ} \eem \quad \ns~,
\end{equation}
where the masses $m^{ij}$ are defined by
\begin{equation}\label{weyl.mass}
m^{ij}=h(q\F^i-\qm\F^j)~.
\end{equation}
The mass matrix in the Cartan subalgebra is symmetric, but in general not 
diagonal,  we find
\begin{equation}
\cMh^{IJ}=h(q-\qm)\f^K\trF\left(H^IH^JH^K\right)=\cMh^{JI}~.
\end{equation}
In the limit of vanishing deformation the above expression is obviously zero,
and we will denote the limit of the masses in \eqn{weyl.mass} by
\begin{equation}
m^{ij}\underset{h=g}{\xrightarrow{q=1}}m_0^{ij}=g(\F^i-\F^j)~.
\end{equation}
It is now straightforward to calculate the mass squared matrix, 
it is also block diagonal and has the non-zero components
\begin{subequations}\label{mass2}\begin{align}
\left(\cMh^\dag\cMh\right)^{ijkl}&=\left(\cMh\cMh^\dag\right)^{ijkl}
=|m^{ki}|^2\d_{il}\d_{jk} \label{weight.mass2} \\ \label{cartan.mass2}
\left(\cMh^\dag\cMh\right)^{IJ}&=\left(\cMh\cMh^\dag\right)^{JI}
=|h(q-\qm)|^2\fb^L\f^M\h^{IJLM}~,
\end{align}\end{subequations}
where
\begin{equation}\label{eta.defn} \h^{IJLM}=
\trF\left(H^IH^JH^LH^M\right)-\frac1N\d^{IL}\d^{JM}~.
\end{equation}

To proceed in the one and two-loop calculations below, we will need to 
assume that the eigenvalues and eigenvectors of the mass squared matrix are
known, that is, we know a unitary matrix $U$ such that
\begin{equation}\label{mass.eigensystem}
\left(U^\dag\cMh^\dag\cMh U\right)^{IJ}=|m_I|^2\d^{IJ} \quad \ns~.
\end{equation}
We also need the trace of the mass squared operator.  This requires the
trace of $\h^{IJLM}$, which can be found using the
completeness relation for the Cartan subalgebra.  The final expression
is simplified by using the tracelessness of $\F$ to get
\begin{align}\label{mass.trace}\non
\tr|\cMh|^2&=|h|^2\left(\sum_{i\neq j}|m^{ij}|^2+\sum_I|m_I|^2\right)\\
&=N|h|^2\left(|q|^2+|\qm|^2-\frac2{N^2}|q-\qm|^2\right)\tr\F^\dag\F~.
\end{align}
\section{One-loop K\"ahler potential}\label{1loop.sect}
From the quadratic terms defined in \eqn{quad.class.act} one can 
read off, see e.g. \citep{Kuzenko:2005CS}, the one-loop effective action as
\begin{equation}\label{oneloop.ea.1} \G^{(1)}=
\rmi\Tr\ln((\Box-|\cMh|^2)P_+)-\rmi\Tr\ln((\Box-|\cMg|^2)P_+)~,
\end{equation}
where $\Tr$ is both a matrix trace and a trace over full superspace, and
$P_+=(16\Box)^{-1}\Db^2D^2$ is the (flat) chiral projection operator.  
The matrix trace can be converted to a sum of eigenvalues using the results of 
section \ref{CWbMO.sect}. 
The functional trace reduces to calculating the standard momentum integral 
(in $d=4-2\eps$ dimensions)
\begin{equation}
\scJ(m^2)=-\rmi  \int\frac{\m^{4-d}\rmd^dk}{(2\p)^d}
\frac1{k^2}\log\left(1+\frac{m^2}{k^2}\right)
=\frac{m^2}{(4\p)^2}\left(\k_M-\log\frac{m^2}{M^2}\right)~,
\end{equation}
where $M$ is an arbitrary mass scale, and 
$\k_M=\frac1\eps+2-\log\frac{M^2}{\bar\m^2}+\ord(\eps)$ 
with the $\overline{MS}$ renormalisation point defined by 
$\bar\m^2=4\p\m^2\rme^{-\g}$.
The function $\k_M$ contains all of the information about the 
method of regularisation, e.g. if we had regularised by using a 
momentum cut-off at $\L^2$
then we would have had $\k_M=1-\log\frac{M^2}{\L^2}$~.

Factoring out the integral over full superspace we get the 
one-loop K\"ahler potential
\begin{align}\label{1loop.eqn1}K^{(1)}&=\sum_I\scJ(|m_I|^2)
  +\sum_{i\neq j}\left(\scJ(|m^{ij}|^2)-\scJ(|m_0^{ij}|^2)\right)
\end{align}
As a check on the above result, we note that it is zero in the limit of 
vanishing deformation. 
The $\k_M$-dependent terms are proportional to the trace of the difference 
of the deformed and undeformed mass matrix
\begin{equation} 
\sum_I|m_I|^2+\sum_{i\neq j}\left(|m^{ij}|^2-|m_0^{ij}|^2\right)=
\trad\left(|\cMh|^2-|\cMg|^2\right) ~.
\end{equation}
So using the trace formula \eqn{mass.trace} it is easily seen that the above
term is zero if the well known one-loop finiteness condition holds 
\citep{Freedman:2005cg,Penati:2005hp,Rossi:2005mr},
\begin{equation}\label{finite.cond}
2g^2=|h|^2\left(|q|^2+|\qm|^2-\frac2{N^2}|q-\qm|^2\right)\equiv2f_q|h|^2~.
\end{equation}

If we enforce the finiteness condition and choose $M^2$ to be any 
field generated mass then we get the explicitly super-conformal result
\begin{align}
(4\p)^2K^{(1)}&=
\sum_{i\neq j}\left(|m_0^{ij}|^2\log\frac{|m_0^{ij}|^2}{M^2}
  -|m^{ij}|^2\log\frac{|m^{ij}|^2}{M^2}\right)
-\sum_I|m_I|^2\log\frac{|m_I|^2}{M^2}~.
\end{align}
We emphasise that this result is independent of the choice of $M^2$.
\section{Two-loop K\"ahler potential}\label{2loop.sect}
\begin{figure}[t]
  \centering
  \hfill\scalebox{0.8}{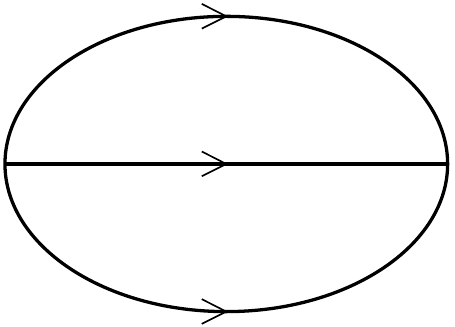}
  \hfill\scalebox{0.8}{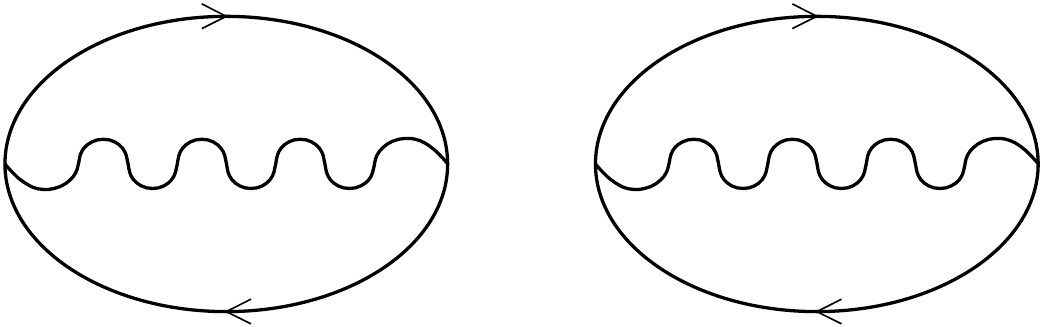}\hspace{20mm}
  \caption{The two-loop diagrams contributing to the K\"ahler potential,
	$\G_1$ and $\G_2$ respectively.
	The arrows show the flow of chirality around the loop, while the
	fields label the propagators.  The squiggly line corresponds to 
	the $\cN\!=\!1$ gauge superfield.}
  \label{2loop.fig}
\end{figure}
In the $\b$-deformed theory there are only four two-loop diagrams that differ 
from the undeformed theory \citep{Kuzenko:2007tf}, 
but some simple D-algebra shows that only two give non-zero contributions 
to the K\"ahler potential (see Fig. \ref{2loop.fig}).
Both are of the sunset type and have the generic group theoretic structure
\begin{equation}
\G=\k\intz\rmd^8z'G^{ab}\trad\left(
T^a_{(h,\qm)}\hat G_{(h,q)}T^b_{(\hb,\qbm)}\check{G}'_{(h,q)}\right)~,
\end{equation}
where $G$ is an undeformed Green's function and, in general,
$\hat G$ and $\check G$ denote spinor derivatives of deformed Green's functions.
This decomposes in the Cartan-Weyl basis into three terms,
\begin{align}\label{decomp.CW} \G=&{}\k |h|^2 \intz\rmd^8z'\Big(
	G^{ijji}\big(q\qb\hat{G}^{j kkj}_{(h,q)}\check{G}'^{ikki}_{(h,q)}
	 +(q\qb)^{-1}\hat{G}^{k iik}_{(h,q)}\check{G}'^{kjjk}_{(h,q)}\big)\non\\
	&+(q(H_K)_{jj}-q^{-1}(H_K)_{ii})(\qb(H_L)_{jj}-\qb^{-1}(H_L)_{ii})\times
	\non\\\non&\qquad\qquad\qquad\times
	\big(G^{KL}\hat{G}^{ijji}_{(h,q)}\check{G}'^{ijji}_{(h,q)}
	+ G^{jiij}\hat{G}^{KL}_{(h,q)}\check{G}'^{jiij}_{(h,q)}
	+ G^{ijji}\hat{G}^{jiij}_{(h,q)}\check{G}'^{LK}_{(h,q)}\big)\Big.\\
   &+ |q-q^{-1}|^2 G^{IJ}\hat{G}^{MN}_{(h,q)}\check{G}'^{LK}_{(h,q)}
      \trF(H_IH_KH_M)\trF(H_JH_LH_N) \Big) \\\non
   =&{} \G_A+\G_B+\G_C ~.
\end{align}
We should note that if the vertices are undeformed, \ie 
$T^a_{(h,q)}\to T^a_{(g,1)}=gT^a_{\rm ad}$, 
then the final term, $\G_C$, is zero.

For an arbitrary background in the Cartan subalgebra
$\G_A$ is easy to evaluate as all of its Green's functions are diagonal.  
To evaluate the other terms, which involve sums over the Cartan subalgebra,
we will use the unitary matrices defined in 
\eqn{mass.eigensystem} to diagonalise the Green's functions, 
\begin{equation}
(H_I)_{jj} \Ghl_{(h,q)}^{IJ} (H_J)_{ii}
=(H_I)_{ii} \Ghr_{(h,q)}^{IJ} (H_J)_{jj}
=(\bar\bH_K)_{ii}G_{(h,q)}^{(K)}(\bH_K)_{jj}~.
\end{equation}
The modified diagonal generators are defined by
\begin{equation}
\bH_I=U_I^{~J}H_J~,\quad \bar\bH_I=H_J (U^\dag)^J_{~I}~.
\end{equation}
In the next subsection, these generators will be combined into coefficients 
for the scalar loop integrals.  Alternatively we could 
reabsorb the diagonalising unitary matrices
back into the loop integrals to get a matrix valued expression.
Although this does make some expressions look a bit neater and keep
all of the field dependence in the now matrix valued loop integrals, 
to evaluate the these expressions we would still have to diagonalise the 
mass matrices. 

\subsection{Evaluation of \texorpdfstring{$\G_\I$}{Gamma 1}}
The first diagram we evaluate has the analytic expression
\begin{align}\label{gammaI} \G_{\I} 
  &=\!-\frac1{2^8}\intz\rmd^8z'G^{ab}_{(g,1)}(z,z')\trad\Big(
    T_{(h,\qm)}^a\Db^2D^2\Ghl_{(h,q)}(z,z')
    T_{(\hb,\qbm)}^bD'^2\Db'^2\Ghr_{(h,q)}(z',z)\Big)~.
\end{align}
For a nonzero result to occur when integrating over $\rmd^4\q'$, 
all chiral derivatives have to hit the Grassmann delta functions contained in 
the deformed propagators.  Then, writing $\scG$ for the remaining bosonic parts
of the propagators, shifting the $x'$ integration variable to $\r=x-x'$
and using \eqn{decomp.CW} we obtain
\begin{align}
K_{\I}	=&-|h|^2\intrho\Big\{
\Gb^{ijji}_{(g,1)}\big(q\qb{\Gb}^{jkkj}_{(h,q)}{\Gb}'^{ikki}_{(h,q)}
	+(q\qb)^{-1}{\Gb}^{kiik}_{(h,q)}{\Gb}'^{kjjk}_{(h,q)}\big)\non\\
&+(q(H_K)_{jj}-q^{-1}(H_K)_{ii})(\qb(H_L)_{jj}-\qb^{-1}(H_L)_{ii})\times\Big.
	\non\\\non&\qquad\qquad\qquad\times
	\big(\Gb^{KL}_{(g,1)}\Gb^{ijji}_{(h,q)}\Gb'^{ijji}_{(h,q)}
	+\Gb^{jiij}_{(g,1)}\Gbhl^{KL}_{(h,q)}\Gb'^{jiij}_{(h,q)}
	+\Gb^{ijji}_{(g,1)}\Gb^{jiij}_{(h,q)}\Gbhl'^{KL}_{(h,q)}\big)\\
&+ |q-q^{-1}|^2 \Gb^{IJ}_{(g,1)}\Gbhl^{MN}_{(h,q)}\Gbhr'^{LK}_{(h,q)}
      \trF(H_IH_KH_M)\trF(H_JH_LH_N) \Big\} \non\\
=&{} K_{\I A}+K_{\I B}+K_{\I C}~.
\end{align}
By using the symmetries of the propagators, we already made some 
simplifications in the above expression.

Now, as all of the propagators in $K_{\I A}$ are already diagonal, we can move
straight to momentum space and perform the $\r$ integral to get
\[\begin{split} K_{\I A}=
-|h|^2\sum_{i\neq j\neq k}\int\frac{\rmd^dk\rmd^dp}{(2\p)^2d}
 &\frac1{k^2+|m_0^{ij}|^2}\left(
     |q|^2\frac1{p^2+|m^{kj}|^2}\frac1{(k+p)^2+|m^{ki}|^2} \right.\\
  &\left.+|\qm|^2\frac1{p^2+|m^{ik}|^2}\frac1{(k+p)^2+|m^{jk}|^2}\right)~.
\end{split}\]
Then, using the results and notation of the appendix we have
\begin{equation} K_{\I A}=
 |h|^2\sum_{i\neq j\neq k}\left(
 |q|^2I(|m_0^{ij}|^2,|m^{ki}|^2,|m^{kj}|^2)
+|\qm|^2I(|m_0^{ij}|^2,|m^{ik}|^2,|m^{jk}|^2)\right)~.
\end{equation}

To evaluate $K_{\I B}$ we diagonalise the propagators, as described above.
The result is
\begin{align}
K_{\I B}=&-|h|^2\sum_{i\neq j,K}\intrho
	(q(\bar{\bH}_K)_{jj}-q^{-1}(\bar{\bH}_K)_{ii})
	(\qb({\bH}_K)_{jj}-\qb^{-1}({\bH}_K)_{ii})\times
	\non\\&\qquad\qquad\qquad\times
	\big(\Gb^{(K)}_{(g,1)}\Gb^{ijji}_{(h,q)}\Gb'^{ijji}_{(h,q)}
	+\Gb^{jiij}_{(g,1)}\Gb^{(K)}_{(h,q)}\Gb'^{jiij}_{(h,q)}
	+\Gb^{ijji}_{(g,1)}\Gb^{jiij}_{(h,q)}\Gb'^{(K)}_{(h,q)}\big)\\\non	
	=&{} |h|^2 \sum_{i\neq j, K}\varpi_{\qb,Kij}\left(~
	 I(\,0\,,|m^{ji}|^2,|m^{ji}|^2) 
	 + 2 I(|m_0^{ij}|^2,|m^{ij}|^2,|m_K|^2)
	  \right) ~,
\end{align}
where $\varpi_{q,Kij}$ is defined by
\begin{equation}
\varpi_{q,Kij}=\left(q(\bH_K)_{ii}-\qm(\bH_K)_{jj}\right)
	\left(\qb(\bar\bH_K)_{ii}-\qbm(\bar\bH_K)_{jj}\right)~,\quad\ns~.
\end{equation}

Similarly we evaluate $K_{\I C}$ to find
\begin{align} 
	K_{\I C}=|h(q-\qm)|^2\sum_{I,J}\bm\h_{IJ} I(0,|m_I|^2,|m_J|^2)  ~,
\end{align}
where $\bm\h$ is closely related to $\h$, defined in \eqn{eta.defn},
\begin{equation}
\bm\h_{IJ}=\trF(\bar{\bH}_I\bar{\bH}_J{\bH}_I{\bH}_J )
	-\frac1N(U^TU)_{IJ}(U^\dag U^*)_{JI}~,\quad\ns~.
\end{equation}

Note that in general the coefficients $\varpi_{q,Kij}$ and $\bm\h_{IJ}$ are
functions of ratios of the background dependent masses.
\subsection{Evaluation of \texorpdfstring{$\G_\II$}{Gamma 2}}
The second diagram,
\begin{equation}\label{gammaII}\begin{split} \G_\II 
  &=\frac1{2^9}\intz\rmd^8z'G^{ab}_{(g,1)}(z,z')\trad\Big(
    T_{(g,1)}^a\Db^2D^2\Ghl_{(h,q)}(z,z')
    T_{(g,1)}^bD'^2\Db'^2\Ghl_{(h,q)}(z',z)
  \Big.\\&\Big.\qquad+
  T_{(g,1)}^a\Db^2D^2\Ghr_{(h,q)}(z,z')
    T_{(g,1)}^bD'^2\Db'^2\Ghr_{(h,q)}(z',z)\Big)~,
\end{split}\end{equation}
is simpler due to the lack of deformed vertices.  Following the same procedure 
as above we find $K_{\II}=K_{\II A}+K_{\II B}$, with
\begin{equation}
K_{\II A}=-g^2 \sum_{i\neq j\neq k}\left(~
	 I(|m_0^{ji}|^2,|m^{kj}|^2,|m^{ki}|^2)+
	 I(|m_0^{ji}|^2,|m^{ik}|^2,|m^{jk}|^2)~\right)
\end{equation}
and
\begin{equation}
K_{\II B}=-g^2 \sum_{i\neq j, K}\varpi_{1,Kij}\left(~
	 I(\,0\,,|m^{ji}|^2,|m^{ji}|^2) 
	 + 2 I(|m_0^{ij}|^2,|m^{ij}|^2),|m_K|^2\right)~.
\end{equation}
\subsection{Finiteness and conformal invariance}
Combining the two diagrams we see that, like the one-loop, the two-loop 
K\"ahler potential is written as the difference 
of terms that cancel in the limit of vanishing deformation:
\begin{align}\label{K2.all} K^{(2)}=&\sum_{i\neq j\neq k}
   \big((|hq|^2-g^2) ~ I(|m_0^{ij}|^2,|m^{ki}|^2,|m^{kj}|^2)
	 +(|h\qm|^2-g^2)I(|m_0^{ij}|^2,|m^{ik}|^2,|m^{jk}|^2)\big) \non\\\non
	 &+\sum_{i\neq j, K}\left(|h|^2\varpi_{\qb,Kij}-g^2\varpi_{1,Kij}\right)
	 \left(\,I(0,|m^{ji}|^2,|m^{ji}|^2) 
	 + 2 I(|m_0^{ij}|^2,|m^{ij}|^2,|m_K|^2)\right)\\
	 &+|h(q-\qm)|^2\sum_{I,J}\bm\h_{IJ} I(0,|m_I|^2,|m_J|^2)~.
\end{align}
As described in the appendix, the two-loop integral, $I(x,y,z)$, can be 
decomposed as
\begin{equation} I(x,y,z)=\i(x)+\i(y)+\i(z)+\dsI(x,y,z)~,
\quad \dsI(x,y,z)=-\half\tilde\x(x,y,z)~,
\end{equation}
where the $\i$ terms include all of the divergences and renormalisation point
dependence, and $\tilde\x$, defined in \eqn{xi-tilde}, is a 
function of mass ratios only.
Since the masses are disentangled in the $\i$ terms, 
the sums can be simplified by using the following identities\footnote{
Note that using \eqn{varpisum} it becomes possible to perform the sum over $K$
in the first term of the middle line of \eqn{K2.all}.}:
\begin{subequations}\label{sums}\begin{align}
\label{varpisum} \sum_K\varpi_{q,Kij}
  &=\frac{1}{N}\sum_{i\neq j}\varpi_{q,Kij}
	= |q|^2+\frac1{|q|^2}-\frac1N|q-\qm|^2 :=2g_q \\
\label{etasumJ}	\sum_J\bm\h_{IJ}
	&=\sum_J\h_{IIJJ}=\frac{N-2}N~.
\end{align}\end{subequations}
The result is that all $\i$ dependence can be collected into
\begin{equation}\begin{split} K_\i^{(2)}=&
N\left(|h|^2\left(|q|^2+|\qm|^2-\frac2{N^2}|q-\qm|^2\right)-2g^2\right)\\&%
\times\left(\sum_{i\neq j}\left(\i(|m_0^{ij}|^2)+2\i(|m^{ij}|^2)\right)
+2\sum_{I}\i(|m_I|^2)\right)~.
\end{split}\end{equation}

From the above expression and the trace formulae given in section 
\ref{CWbMO.sect} we may read off the quadratic dependence of the 
K\"ahler potential:
\[
K^{(2)}_{\rm quad}\propto 4N^2(|h|^2f_q-g^2)(2|h|^2f_q+g^2)\tr\F^\dag\F~,
\] 
where the constant of proportionality is a number that is 
subtraction scheme dependent and
$f_q$ is the function 
that occurs in the one-loop finiteness condition \eqn{finite.cond}.
The above prefactor is, for good reason, reminiscent of the general
expression for the two-loop anomalous dimension given in, for example, 
\citep{West:1984dg,Parkes:1985hj,Jack:1996qq,Hamidi:1984ft}.

So, as expected, the two-loop K\"ahler potential is finite and independent of
the renormalisation point if the one-loop finiteness condition, 
\eqn{finite.cond}, is satisfied.  
It is interesting to note that the `meaning'
of \eqn{finite.cond} is different at one and two-loops.  At one-loop it implies
that the trace of the mass matrix is invariant under the deformation, 
while at two loops it implies that the coefficients of the scalar 
diagrams sum to zero.

If we enforce the finiteness condition, \eqn{finite.cond}, then we get the
explicitly superconformal two-loop K\"ahler potential by making the replacements
$g^2\to|h|^2f_q$ and $I\to\dsI$ in \eqn{K2.all}.

\section{Special backgrounds}\label{simpBG.sect}
In the above analysis the background superfield pointed in an arbitrary 
direction in the Cartan subalgebra of $SU(N)$.
In order to make our previous analysis concrete 
we now choose the specific background
\begin{equation}\label{background}
\F=\sqrt{N(N-1)}\f_1H_{N-1}+\sqrt{(N-1)(N-2)}\f_2H_{N-2}~.
\end{equation}
The characteristic feature of this background is that it leaves the 
subgroup $U(1)^2\times SU(N-2)$ of $SU(N)$ unbroken.  
The two $U(1)$s are associated with the generators $H_{N-1}$
and $H_{N-2}$.  In the limit $\f_2\to0$, we obtain the background previously
used for the calculation of the two-loop K\"ahler potential in 
\citep{Kuzenko:2007tf}. 

There are twelve different, nonzero masses that occur with this background.
There are nine deformed masses:
\begin{subequations}
\begin{align}
m_1^2&=|m^{ij}|^2=|m_I|^2=|h(q-\qm)|^2|\f_1+\f_2|^2~,&& \non\\
m_2^2&=|m^{i(N-1)}|^2=|h(q-\qm)\f_1+h(q+(N-2)\qm)\f_2|^2~, 
 &m_{\tilde 2}^2&=|m^{(N-1)j}|^2=m_2^2\big|_{q\to\qm}~, \non\\
m_3^2&=|m^{iN}|^2=|hq\f_2+h(q+(N-1)\qm)\f_1|^2~,
 &m_{\tilde 3}^2&=|m^{Nj}|^2=m_3^2\big|_{q\to\qm}~, \non\\
m_4^2&=|m^{(N-1)N}|^2=|h(N-2)q\f_2-h(q+(N-1)\qm)\f_1|^2~,
 &m_{\tilde 4}^2&=|m^{N(N-1)}|^2=m_4^2\big|_{q\to\qm}~, \non\\
m_\pm^2&=\half|h(q-\qm)|^2\left(a+c\pm\sqrt{(a-c)^2+4|b|^2}\right)~,
\end{align}
where the indices $i,j$ and $I$ range from $1$ to $(N-2)$ and $(N-3)$ 
respectively, and their three undeformed counterparts:
\begin{align}
m_{02}^2&=|m_0^{i(N-1)}|^2=|m_0^{(N-1)j}|^2=g^2(N-1)^2|\f_2|^2~, \non\\
m_{03}^2&=|m_0^{iN}|^2=|m_0^{Nj}|^2=g^2|N\f_1+\f_2|^2~, \non\\
m_{04}^2&=|m_0^{(N-1)N}|^2=|m_0^{N(N-1)}|^2=g^2|N\f_1-(N-2)\f_2|^2~.
\end{align}\end{subequations}
The quantities $a$, $b$ and $c$ come from the Cartan subalgebra block of the 
mass matrix, which is diagonal except for the bottom $2\times2$ block: 
\begin{subequations}
\begin{align}\label{Cartan.mass.bg}
 &(\cMh^\dag\cMh)_{IJ}=|h(q-\qm)|^2
\left(\begin{array}{ccc;{1pt/3pt}cc}
	|\f_1+\f_2|^2 & & & &\\ 
	& \ddots & & & \\
	& & |\f_1+\f_2|^2 & & \\\hdashline[1pt/3pt]
	& & & a & b \\
	& & & b^* & c	 
	\end{array}\right)~, \\
&a=\left((N-3)^2+1-2/N\right)\fb_2\f_2+\fb_1\f_1-(N-3)(\fb_2\f_1+\fb_1\f_2)\\
&b=\sqrt{1-2/N}\left((3-N)\fb_2\f_2+\fb_1\f_2-(N-2)\fb_2\f_1\right)~,\\
&c=(N-2)^2\fb_1\f_1+\big(1-2/N\big) \fb_2\f_2~.
\end{align}
\end{subequations}
The eigenvalues are $m_1^2$ and $m_\pm^2$ with the corresponding
orthonormal eigenvectors
\begin{equation}
e_{I<N-2}~, \quad
v_\pm=\frac{(0,\ldots,0,a-c\pm\s,2b^*)}{\sqrt{
2\s(\s\pm(a-c))}}~,\quad
\s=\sqrt{(a-c)^2+4|b|^2}
\end{equation}
where $e_I$ is the standard basis vector 
with a one in the $I^{\rm th}$ position and zero everywhere else.   
Note that \eqn{Cartan.mass.bg} is diagonal when $\f_2=0$ 
(including the $SU(2)$ case) and in the planar limit, when $N\to\infty$.

The one-loop K\"ahler potential is simply read from \eqn{1loop.eqn1}:
\begin{align}\label{1loop.bg1}
K^{(1)}=&{}\scJ(m_+^2)+\scJ(m_-^2)-
2(N-2)\big(\scJ(m_{02}^2)+\scJ(m_{03}^2)\big)-2\scJ(m_{04}^2)
+(N^2-2N-1)\scJ(m_1^2)\non\\
&+(N-2)\big(\scJ(m_2^2)+\scJ(m_{\tilde2}^2)
+\scJ(m_3^2)+\scJ(m_{\tilde3}^2)\big)+\scJ(m_4^2)+\scJ(m_{\tilde4}^2)~.
\end{align}
The effect of enforcing the finiteness condition 
is to replace $(4\p)^2\scJ(x)$ by $x\log(M^2/x)$ 
for an arbitrary field dependent mass term $M^2$.
Similarly, the two-loop K\"ahler potential is read from \eqn{K2.all}:
\begin{align}\label{K2.2dbg} K^{(2)}=&{} \non
(N-2)\big(|hq|^2-g^2\big)\Big[
 (N-3)(N-4)I(0,1,1)+2I(2_0,\tilde3,\tilde4)+2I(3_0,\tilde2,4)+
 2I(4_0,2,3)\\\non &\quad+(N-3)
 \big(2I(2_0,2,1)+2I(3_0,3,1)+I(0,\tilde2,\tilde2)+I(0,\tilde3,\tilde3)\big)
 \Big]+\big[q\to\qm\big] \\\non
&+(|h|^2g_q-g^2)\Big[(N-2)\Big((N-3)I(0,1,1)\\\non
  &\quad+2I(0,2,2)+2I(0,3,3)\Big)+2I(0,4,4)\Big]+\big[q\to\qm\big]\\\non
&+2\sum_{i\neq j}\Big[ \big(2|h|^2g_q-2g^2-\vpp_{\qb,2ij}-\vpp_{\qb,1ij}\big)
 I(|m^{ij}_0|^2,|m^{ij}|^2,1)\\
&\quad+\vpp_{\qb,2ij}I(|m^{ij}_0|^2,|m^{ij}|^2,+)
 +\vpp_{\qb,1ij}I(|m^{ij}_0|^2,|m^{ij}|^2,-)
 \Big]
\\\non&+|h(q-\qm)|^2\Big[
 \big((1-2/N)(N-5)+\hp_{22}+2\hp_{21}+\hp_{11}\big)I(0,1,1)\\\non
&\quad+2(1-2/N-\hp_{22}-\hp_{21})I(0,+,1)
 +2(1-2/N-\hp_{12}-\hp_{11})I(0,-,1) \\\non
&\quad+\hp_{22}I(0,+,+)+2\hp_{21}I(0,-,+)+\hp_{11}I(0,-,-)\Big],
\end{align}
where we have introduced a condensed notation for the masses
\[ m^2_\pm\sim\pm~,\quad m^2_i\sim i \quad\mbox{and}\quad m^2_{0i}\sim i_0 \]
with $i=1,\tilde1,\ldots,4,\tilde4$ and defined
\[ \vpp_{q,Kij}=|h|^2\varpi_{q,(N-K)ij}-g^2\varpi_{1,(N-K)ij}~,
\quad \hp_{IJ}=\bm\h_{(N-I)(N-J)}~.
\]
We've also used \eqn{sums} to make the expression only dependent
on $\vpp_{q,Iij}$ and $\hp_{IJ}$ for $I,J=1,2$.
The coefficients, $\varpi_{q,Kij}$ and $\bm\h_{IJ}$, 
are then calculated using the results
\begin{align*} \bH_I&=H_I~,\quad I<N-2~,\\
\quad \bH_{N-2}&=(v_+)_{N-2}H_{N-2}+(v_-)_{N-2}H_{N-1}~,\quad
\quad \bH_{N-1}=(v_+)_{N-1}H_{N-2}+(v_-)_{N-1}H_{N-1}~.
\end{align*}
We emphasise that $\bH_I$ and therefore $\varpi_{q,Kij}$ and $\bm\h_{IJ}$ 
are in general field dependent quantities.

We now examine the two limiting cases, $\f_2\to0$ and $N\to3$.  
In both these limits we find that
the coefficients $\vpp_{q,Iij}$ and $\hp_{IJ}$ are independent of the background
fields, which is not representative of the general case.

In the case where $\f_2\to0$ the entire mass matrix is diagonal, so that
the unitary, diagonalising matrix is just the unit matrix.
Thus the coefficients $\varpi$ and $\bm\h$ are background independent, 
and can be calculated in closed form. 
Also, similarly to \eqn{K2.2dbg}, we can write
$K^{(2)}$ such that we only need to know $\bm\h_{(N-1)(N-1)}$ and 
$\varpi_{q,(N-1)ij}$, which further eases the calculational load.
If we enforce conformal invariance then the sole mass scale, $\fb_1\f_1$,
must cancel in all of the mass ratios, so that the full, quantum corrected, 
K\"ahler potential is just a deformation dependent
rescaling of the classical K\"ahler potential \citep{Kuzenko:2007tf}.
Finally, if we choose a real deformation,
the limit of our two-loop result reproduces equation (6.5) of 
\citep{Kuzenko:2007tf} exactly, which is a good check of our method.

When the gauge group is $SU(3)$ the terms with the mass $m_1^2$ 
no longer appear in the summations, $m_\pm^2$ is compactly written as 
$|h(q-\qm)(\f_1\mp\frac\rmi{\sqrt3}\f_2)|^2$ and the 
rest of the masses take the obvious limits. 
We will assume that we are on the conformal surface and set
$g^2=f_q|h|^2$.
The one-loop K\"ahler potential does not simplify much, choosing 
$M^2=f_q|h\f|^2$ where $|\f|^2=\tr\F^\dag\F\neq0$, we have
\begin{align}
\frac{(4\p)^2}{|h|^2}K^{(1)}_{SU(3)}
=&{}2f_q\left(\Maz\log\frac\Maz{|\f|^2}+\Mbz\log\frac\Mbz{|\f|^2}
	+(\f_2\to-\f_2)\right) \non\\\non
&-|q-\qm|^2\left(|\f_1-\frac\rmi{\sqrt3}\f_2|\log\frac{|\f_1-\rmi\f_2/\sqrt3\,|}{f_q|q-\qm|^{-2}|\f|^2}
	+(\f_2\to-\f_2)\right)\\\non
&\hspace{-30pt}-\Bigg(\Ma\log\frac\Ma{f_q|\f|^2}\\\non
&\hspace{-20pt}+\Mb\log\frac\Mb{f_q|\f|^2}+(\f_2\to-\f_2)\Bigg)
+\left(q\to\qm\right)~.
\end{align}
Although we can combine the logarithms and explicitly remove all reference to
$|\f|^2$, the analytic structure and the various limits are simpler
to examine in the above form.

To find the two-loop K\"ahler potential, we choose
the diagonalising unitary matrix to be
\[ U=\frac1{\sqrt2}\bem 1 & \rmi \\ -\rmi & -1\eem 
\quad \implies \bH_2=(\rmi\bH_1)^*=\frac{-1}{\sqrt3}\diag(r_+,r_-,-1) 
\]
where $-1$ and $r_\pm=\half(1\pm\rmi\sqrt3)$ are the cube roots of minus one.
Then it is straightforward to compute
\[ \varpi_{q,Kij}=\frac13\left[\bem 
|q-\qm|^2 & |qr_--\qm r_+|^2 & |qr_-+\qm|^2
\\|qr_+-\qm r_-|^2 & |q-\qm|^2 & |qr_++\qm|^2
\\|q+\qm r_-|^2 & |q+\qm r_+|^2 & |q-\qm|^2
\eem ,q\to \qm \right]~,
\quad \bm\h_{IJ} = \frac13\d_{IJ} ~. \]
We split the two-loop K\"ahler potential into
$K_{SU(3)}^{(2)}=K_{\rm A}+(q\to\qm)+K_{\rm B}+(q\to\qm)$
where the labelling follows the decomposition \eqn{decomp.CW}.
Note that in the case being considered
$K_{C}=0$, since it only contributes terms of the
form $\dsI(0,x,x)$ which are zero from \eqn{xi-tilde}.
This is also true for the integrals that come from the first terms in 
$K_{\I B}$ and $K_{\II B}$.
Substituting in the masses and using the fact that $\dsI(x,y,z)$ 
is a homogeneous function of order one to pull out a factor of $|h|^2$,
we find
\begin{align*} 
K_{\rm A}^{(2)}=&{} 2|h|^4(|q|^2-f_q)\Big[
 \dsI\big(\MaZ,\Mbt,\Mct\big)\\
&+\dsI\big(\MbZ,\Mat,\Mc\big)\\
&+\dsI\big(\McZ,\Ma,\Mb\big) \Big]
\end{align*}
and
\begin{align*} 
K_{\rm B}^{(2)}=&{}2|h|^4\Bigg[ \big(\third|qr_--\qm r_+|^2-f_q\big)\\
&\hspace{-0 pt}\times\Big[\dsI\Big(\MaZ,\Ma,\Mp\Big)\\
	&\quad+\dsI\Big(\MaZ,\Mat,\Mm\Big)\Big]\\
&\hspace{-25pt}+\big(\third|q r_-+\qm|^2-f_q\big)\Big[\dsI\Big(\MbZ,\Mb,\Mp\Big)\\
	&\quad+\dsI\Big(\MbZ,\Mbt,\Mm\Big)\Big]\\
&\hspace{-25pt}+\big(\third|qr_+ +\qm|^2-f_q\big)\Big[\dsI\Big(\McZ,\Mc,\Mp\Big)\\
	&\quad+\dsI\Big(\McZ,\Mct,\Mm\Big)\Big] \Bigg]
\end{align*}
From the expressions for $\dsI=-\tilde\x/2$ given in the appendix, 
\eqn{xi-tilde} and \eqn{homog-xi},
we see that the above form is scale invariant. 
We note that taking the deformation to be real does not provide much
simplification, except when $\f_2=0$ and a real deformation
makes the tilded masses equal to their non-tilded counterparts.

\section{Conclusion}\label{conc.sect}
The above calculations show that although it is conceptually straightforward
to calculate the loop corrections to the K\"ahler potential of $\b$-deformed
$\cN\!=\!4$ SYM on the Coulomb branch, 
the details of the calculation are quite involved for an arbitrary background.  
This is because not only do the $\half(3N-2)(N-1)$ eigenmasses enter the 
result, but also the field dependent eigenvectors.

To help reveal the general structure of the K\"ahler potential it is useful
to use the idea of matrix valued loop integrals 
(see e.g. \citep{Nibbelink:2005wc}) discussed in section \ref{2loop.sect}. 
Then all field dependence is in the loop integrals, for example
\[ \sum_{IJ}\bm\h_{IJ}I(0,m_I^2,m_J^2)
 =\sum_{IJKL}\h_{IJKL}I(0,(\cM^\dag\cM)^{IJ},(\cM\cM^\dag)^{KL})~.
\]
Thus we see that, assuming the finiteness condition is enforced,
the general conformally invariant structure of the 
K\"ahler potential can be written in terms of a function of 
the $\half(5N-2)(N-1)$ components of the mass matrix \eqn{mass2}
\begin{align*} K(\F^\dag,\F)=&{}|\f|^2 \, F\left(
\frac{|g(\F^i-\F^j)|^2}{|\f|^2},
\frac{|h(q\F^i-\qm\F^j)|^2}{|\f|^2},
\frac{|h(q-\qm)|^2\fb^L\f^M\h_{IJLM}}{|\f|^2}\right)~,
\end{align*}
where we remember that we have chosen the background to be 
$\F=H_I\f^I=E_{ii}\F^i$.
For definiteness, we have inserted the nonvanishing 
$|\f|^2=\tr\F^\dag\F=\sum_I|\f^I|^2=\sum_i|\F^i|^2$ into all
terms in the above expression, but in general this is not necessary.
The loop corrections to the K\"ahler potential are identically zero in 
the limit of vanishing deformation, thus $F$ can always be written as one 
(for the tree level term) plus the
difference between two terms that become identical as the deformation is 
switched off.

Finally, we should compare our results to the two-loop K\"ahler potential
calculation of \citep{Nibbelink:2005wc}.  They examined the K\"ahler
potential of a general, non-renormalisable $\cN\!=\!1$ theory with the assumption
that the background chiral fields satisfy the classical equations of motion.
This assumption turns out to be suitable for special background configurations
(such as vacuum valleys) but appears to be incompatible with the K\"ahler 
approximation for general backgrounds \citep{Kuzenko:Disc}.  
The point is that the derivatives of the background 
chiral fields are systematically ignored when computing the quantum corrections 
to the K\"ahler potential. Then, for (dynamically) massive fields, enforcing
the equations of motion restricts the background fields to discrete or vanishing
values%
\footnote{Similar problems occur in non-supersymmetric theories when the 
equations of motion are assumed in an effective potential calculation. 
}.
(For example, a simple model where enforcing the equations of motion
will lead to a vanishing K\"ahler potential is 
massive supersymmetric QED%
  \footnote{The two-loop Euler-Heisenberg Lagrangian and 
  the one-loop K\"ahler potential for SQED were studied in 
  \citep{Kuzenko:2007cg}.  
  The one-loop K\"ahler potential was also studied in a general,
  two-parameter $R_\xi$-gauge \citep{Tyler:2param}.
  }.
The classical equations of motion are 
$\f_\pm=\frac1m\Db^2\fb_\mp$ which imply that the background
fields are zero in the K\"ahler approximation.)
With that said,  
there are many interesting theories 
with background configurations
that do not have the above problem, 
e.g. the Coulomb branch of both $\cN\!=\!2$ and $\b$-deformed SYM theories.
A comparison between the final results of \citep{Nibbelink:2005wc} 
and our initial expressions (before going to the Cartan-Weyl basis) 
has been made and it was found that the two results match%
\footnote{Note that the match only occured after an error in
\href{http://arxiv.org/abs/hep-th/0511004}{{\tt hep-th/0511004}}
was corrected. 
}.

\vspace{5mm}
\noindent {\bf Acknowledgements:}\\
I would like to thank Sergei Kuzenko for suggesting this project
and providing continuing guidance.  
I am also grateful for many useful discussions with
Ian McArthur, Gabriele Tartaglino-Mazzucchelli and Paul Abbott.
I thank Stefan Groot Nibbelink and Tino S. Nyawelo for informative 
correspondence regarding effective K\"ahler potentials.
Finally I acknowledge helpful correspondence with Kristian McDonald and 
Stephen Martin regarding two-loop integrals.
This work was supported by an Australian Postgraduate Award.

\begin{appendix}
\section{Closed form for \texorpdfstring{$I(x,y,z)$}{I(x,y,z)}}\label{Ixyz.app}
In this appendix we examine the calculation of the two-loop vacuum diagram
\begin{equation}\label{Ixyz-defn} I(x,y,z)=
  \m^{4\eps}\int\frac{\rmd^dk\rmd^dp}{(2\p)^{2d}}
  \frac1{(k^2+x)(p^2+y)((k+p)^2+z)}
\end{equation}
where we work in a $d=4-2\eps$ dimensional, Euclidean space-time and $x$, $y$
and $z$ are three independent square masses.  In the literature there are four
main approaches to calculating this integral.  It can be directly calculated, as
in \citep{Davydychev:1992mt} where the Mellin-Barnes representation for the
propagators is used, or it can be calculated indirectly by exploiting the 
different types of differential equations \citep{Kotikov:1990kg,Kotikov:1991hm}
that $I(x,y,z)$ has to satisfy.  
The first differential equation is the homogeneity equation,
\begin{equation}\label{homog-eqn}
  (1-2\eps-x\pd_x-y\pd_y-z\pd_z)I(x,y,z)=0~,
\end{equation}
and was used in \citep{Bij:1983bw,Hoogeveen:1985tf,McDonald:2003zj} to express
$I(x,y,z)$ in terms of its first derivatives, which have a more amenable Feynman
parameterisation\footnote{This approach was extended to non-vacuum diagrams in
\citep{Ghinculov:1994sd}.}.
The second type of differential equation is the ordinary differential 
equation of
\citep{Caffo:1998du}\footnote{Also used in \citep{Ford:1991hw} for the case of
two equal masses.}.
The final approach is the partial differential equation used in
\citep{Ford:1992pn}, and is the approach that we'll re-examine here.

Although all approaches must yield equivalent results, only those of
\citep{Caffo:1998du} and \citep{Davydychev:1992mt} had been analytically shown
to be the same (to the authors knowledge).  
This appendix will take this one step further and show
the equivalence of the Clausen function form for $I(x,y,z)$ given in 
\citep{Davydychev:1992mt} to the  result of \citep{Ford:1992pn}
which is expressend in terms of Lobachevsky functions.  
Along the way we find a completely symmetric representation for 
$I(x,y,z)$ that holds for all values of the masses.

By using the integration by parts technique 
\citep{Tkachov:1981wb,Chetyrkin:1981qh}, 
one can see that $I(x,y,z)$ must satisfy the following differential equation
\begin{equation}\label{flow-eqn1}
\left[ (z-y)\pd_x+\cycl\right]I(x,y,z)
 =\left[J'(x)\big(J(y)-J(z)\big)+\cycl\right]~,
\end{equation}
where $J(x)$ is the one-loop tadpole integral
\begin{equation}\label{Jx-defn} J(x)=\m^{2\eps}\intk\frac1{k^2+x}
  =\frac{\m^{2\eps}}{(4\p)^{2-\eps}}\G(\eps-1)x^{1-\eps}~.
\end{equation}
In \citep{Ford:1992pn} it was noted that \eqn{flow-eqn1} can be solved using the 
method of characteristics.  To do this we need to introduce a one-parameter flow
$(x_t,y_t,z_t)$ such that
\begin{equation}\label{mass-flow1}
  \dot x_t=y_t-z_t~,\quad \dot y_t=z_t-x_t~,\quad \dot z_t=x_t-y_t~,
\end{equation}
which allows us to write \eqn{flow-eqn1} as
\begin{equation}\label{flow-eqn1t}
  \ddt{} I(x_t,y_t,z_t)=-\G'\left(\dot x_t(y_tz_t)^{-\eps}
   +\dot y_t(z_tx_t)^{-\eps}+\dot z_t(x_ty_t)^{-\eps}\right)~,
\end{equation}
where $(4\p)^d\G'=\m^{4\eps}\G(\eps)\G(\eps-1)$~. The flow \eqn{mass-flow1} has
two algebraically independent invariants, we choose
\begin{equation}\label{c-Delta-defn}
c=x_t+y_t+z_t~,\qquad \D=2(x_ty_t+y_tz_t+z_tx_t)-x_t^2-y_t^2-z_t^2~,
\end{equation}
where $\D$ is known as the ``triangle'' function
and is related to the negative of the K\"allen function.  
Using these invariants we can write
\begin{equation}
  y_tz_t=\left(x_t-\frac c2\right)^2+\frac\D4~,\quad \mbox{and cycl.}
\end{equation}
allowing us to integrate the flow equation in the form
\begin{equation}I(x,y,z)=I(x_0,y_0,z_0)-\G'
\left(\int_{x_0-c/2}^{x-c/2}+\int_{y_0-c/2}^{y-c/2}+\int_{z_0-c/2}^{z-c/2}
\right)\frac{\rmd s}{(s^2+\D/4)^\eps}~,
\end{equation}
where the end point of the flow has been chosen as $(x_1,y_1,z_1)=(x,y,z)$.
We can now choose the flow's initial point so that the integral 
$I(x_0,y_0,z_0)$ is more easily evaluated.  In \citep{Ford:1992pn} the flow was 
chosen to start at $(X,Y,0)$, a choice that is only good for $\D\leq0$, while in
\citep{Kuzenko:2007tf} the case of $\D>0$ was examined using the initial point
$(X,Y,Y)$.  In this discussion we make the latter choice for all values of $\D$.
The masses $X$ and $Y$ can be seen to be real and non-negative when 
written in terms of the flow invariants using 
\begin{equation}\label{c-Delta-2}
c=x_t+y_t+z_t=X+2Y~,\quad 
\D=2(x_ty_t+y_tz_t+z_tx_t)-x_t^2-y_t^2-z_t^2=X(4Y-X)~.
\end{equation}
Although the explicit form of $I(X,Y,Y)$ is known, we will once again follow
\citep{Ford:1992pn,Kuzenko:2007tf} and make a second flow based on the 
differential equation
\begin{equation}\label{flow-eqn2}
\left(X\pd_X+\left(\half X-Y\right)\pd_Y\right)I(X,Y,Y)
=\G'\frac{X^{1-\eps}-Y^{1-\eps}}{Y^\eps}~,
\end{equation}
which is solved by introducing another flow
\begin{equation}\label{mass-flow2}
\dot X_t=X_t~,\quad \dot Y_t=\half X_t-Y_t~,\quad X_1=X~,\quad Y_1=Y~.
\end{equation}
This flow also conserves the triangle function, $\D$,
but $c$ is no longer preserved.
We now need to choose different starting points for the flow,
depending on the sign of $\D$, specifically we choose
\begin{equation} (X_0,Y_0)=(\sqrt{-\D},0)\quad \mbox{and} \quad
 (X_0,Y_0)=(\sqrt{\D/3},\sqrt{\D/3})
\end{equation}
for $\D<0$ and $\D>0$ respectively.  
The differential equation may now be integrated, yielding
\begin{equation*}
I(x,y,z)=\G'\left[G\left(\frac c2-x\right)+\cycl\right]+
\left\{\begin{aligned}
  I\!\left(\sqrt{-\D},0,0\right)-\G' G\left(\sqrt{-\D/4}\right)~, \quad &\D<0 \\
  I\!\left(\!\sqrt{\frac{\D}{3}},\sqrt{\frac{\D}{3}},\sqrt{\frac{\D}{3}}\right)
	-3\G' G\left(\!\sqrt{\frac{\D}{12}}\right)~, \quad&\D>0~,
\end{aligned}\right.
\end{equation*}
where
\begin{equation}\label{G-defn}
G(w)=\int_0^w\frac{\rmd s}{(s^2+\D/4)^\eps} 
\end{equation}
may be integrated in terms of Gauss hypergeometric functions.
Whereas the diagram $I(x,0,0)$ is easily evaluated using elementary means, the 
equal mass diagram is not so simple.  To proceed we may either use the explicit
form of $I(x,x,x)$ given in the literature, e.g. 
\citep{Broadhurst:1993mw,Tarasov:1997kx}, or analytically continue the result
for $\D<0$. Either way we find
\begin{equation}\label{Ixyz-result}
I(x,y,z)=\sin{\p\eps}\,I\!\left(\sqrt{\D},0,0\right)+
\G'\left[G\left(\frac c2-x\right)+\cycl\right]~,
\end{equation}
a result that holds for arbitrary $\D(x,y,z)$. 
The square root, $\sqrt\D=\exp(\half\log\D)$, 
is always taken on its principle branch.

Finally we examine the expansion of $I(x,y,z)$ around $d=4$.  
This is found by using 
\begin{equation}\label{Ix00} I(x,0,0)=-\frac x{(4\p)^4}
\left(\frac{4\p\m^2}{x}\right)^{2\eps}\G(\eps-1)\G(2\eps-1)\G(1-\eps)
\end{equation}
and expanding the denominator in the integrand of \eqn{G-defn}. The result is
\begin{equation}\label{Ixyz-expn} (4\p)^4 I(x,y,z)
=-\frac{c}{2\eps^2}+\frac{\hat L_1}\eps-\half\left(c\left(\z(2)+\frac52\right)
 +2\hat L_2 +\tilde\x(x,y,z)\right)+\ord(\eps)~,
\end{equation}
where we've used the ``natural'' renormalisation point%
\footnote{
This renormalisation point is only natural when we work with the graph
that has not had its subdivergent graphs subtracted.
}
$\hat\m^2=4\p\m^2\rme^{3/2-\g}$~,
\[ \hat L_n := x\log^n\frac x{\hat\m^2}+y\log^n\frac y{\hat\m^2}
 +z\log^n\frac z{\hat\m^2}~,
\]
and $\tilde\x(x,y,z)$ decomposes as
\begin{align}\label{xi-tilde}
\tilde\x(x,y,z)&=\x(x,y,z)-\left[ x\log(y/x)\log(z/x)+\cycl\right] \\
\x(x,y,z)&=2\sqrt\D\left(N(2\q_x)+N(2\q_y)+N(2\q_z)\right)~.\label{xi}
\end{align}
Note that the above definition of $\x(x,y,z)$ holds for all $x,y,z\geq0$,
in distinction to the separate definitions given in \citep{Ford:1992pn} for 
$\D>0$ and $\D<0$.
In the previous equation we have used the function
\begin{equation}\label{N.defn} N(\q)=
 -\int_0^{\q}\rmd\f\log\left(2\cos\frac\f2\right)~, 
\end{equation}
that is related to the Lobachevsky function 
(and thus the dilogarithm and Clausen function \citep{Lewin:1981}).
It is evaluated on the angles 
\begin{equation}\label{angle.defn}
\q_x=\arctan\frac{-x+y+z}{\sqrt{\D}} \quad \mbox{and cyclic}~. 
\end{equation}
For $\D>0$ the above angles are real and less than $\p/2$, so that
the function \eqn{N.defn} is equivalent to the log-cosine function,
defined through the log-sine function (see e.g. 
\citep{Lewin:1981,Davydychev:2000na}) 
\[ \Lc_j(\q)=\Ls_j(\p)-\Ls_j(\p-\q)~,
\quad \Ls_j(\q)=-\int_0^\q\rmd\f\log^{j-1}\left|2\sin\frac\f2\right|~.
\]
Although $\Ls_2(\q)=\Cl_2(\q)$, it is the log-sine series of functions
rather than the Clausen series that gives the simplest $\eps$-expansion 
for $\D>0$ \citep{Davydychev:1999mq,Davydychev:2000kw}.

The above results may be seen to be equivalent with those in \citep{Ford:1992pn}
by shifting the renormalisation point to $\bar\m^2=\hat\m^2\rme^{-3/2}$ so that
\[ \hat L_2=L_2-3L_1+\frac94 c~,\quad \mbox{with} \quad
 L_n:=x\log^n\frac{x}{\bar\m^2}+\cycl
\]
and then replacing one of the $L_2$'s using
\[ \left[(x-y-z)\log\frac{y}{\bar\m^2}\log\frac{z}{\bar\m^2}+\cycl\right]
 =L_2-\left[ x\log\frac yx\log\frac zx+\cycl\right]~.
\]
Finally the form of $\x(x,y,z)$ can be seen to agree with 
that in \citep{Ford:1992pn} by noting
\[ \q_x+\q_y+\q_z=\sgn(\D)\frac\p2 \]
and,  in the case of $\D<0$, rewriting their $\f_w$ in terms of $\q_w$ using 
the standard formula \citep{functions.wolfram.com}
\[ {\rm arcoth}(z)
 =\rmi\left(\arctan\left(\frac z\rmi\right)\pm\frac\p2\right)~, \quad \pm~
\mbox{if} ~~\pm z>1~. 
\]
It is also straightforward to see, in the case $\D>0$ where
\begin{equation*}
\q_x=\frac\p2-\arccos\frac{-x+y+z}{\sqrt{4yz}} \mbox{ and cyclic}~,
\end{equation*}
that we obtain the result of 
\citep{Davydychev:1992mt,Davydychev:1999mq}\footnote{The latter paper
provides an all order $\eps$-expansion of $I(x,y,z)$ starting from its 
hypergeometric representation (\eqn{Ixyz-result} and 
\citep{Ford:1992pn,Davydychev:1992mt}) and the ``magic connection'' 
\citep{Davydychev:1995mq}.}, 
written in terms of log-sine functions and that our expression is automatically
the correct analytic continuation for $\D<0$\footnote{  
The procedure for analytic continuation advocated in 
\citep{Davydychev:1999mq,Davydychev:2000kw} involves rewriting
the log-sine integrals in terms of the generlised Neilson polylogarithms,
a nontrivial task at higher orders.}. 
It would be interesting to see if rewriting the results of
\citep{Davydychev:1999mq} in terms of the angles \eqn{angle.defn} would
provide the correct analytic continuation to $\D<0$ at all orders in the 
epsilon expansion. 

In conclusion it is interesting to compare the above form of $\tilde\x(x,y,z)$
with that obtained from the method of Veltman and van der Bij
\citep{Bij:1983bw} (see also \citep{Hoogeveen:1985tf,McDonald:2003zj}).  
Their method also leads to a single expression that holds for all 
$x$, $y$ and $z$, but is better at revealing the simple 
mass dependence of $\tilde\x$.  Explicitly we find
\begin{align}
\tilde\x(x,y,z)&=x f(\frac yx,\frac zx)+y f(\frac zy,\frac xy)
  +z f(\frac xz,\frac yz) \label{homog-xi} \\ \label{fab-defn}
f(a,b)&=\int_0^1\rmd\a\left(\Li_2(1-w)+\frac{w\log w}{w-1}\right)~,\quad
 w=\frac a\a+\frac b{1-\a}~,
\end{align}
where the integral in \eqn{fab-defn} can be performed in terms of dilogarithms %
\citep{Bij:1983bw,Hoogeveen:1985tf,McDonald:2003zj}.  We note that $\D$ 
naturally appears during this integration.
\end{appendix}
\providecommand{\href}[2]{#2}\begingroup\raggedright\endgroup
\end{document}

%% file: preamble.tex
\newcommand {\cD}{{\cal D}}

\newcommand {\cM}{{\cal M}}
\newcommand {\cN}{{\cal N}}

\newcommand {\cW}{{\cal W}}

\newcommand{\bH}{{\bf H}}

\newcommand{\dsI}{{\mathbb I}}
\def\ds1{\ensuremath{\mathbbm{1}}}
\def\a{\alpha}
\def\b{\beta}
\def\d{\delta}

\def\f{\phi}
\def\g{\gamma}
\def\G{\Gamma}
\def\h{\eta}
\def\i{\iota}
\def\j{\psi}
\def\k{\kappa}

\def\m{\mu}

\def\p{\pi}
\def\q{\theta}
\def\r{\rho}
\def\s{\sigma}

\def\x{\xi}
\def\z{\zeta}
\def\D{\Delta}
\def\F{\Phi}

\def\L{\Lambda}

\def\S{\Sigma}

\newcommand{\eps}{\varepsilon}
\newcommand{\vf}{\varphi}
\newcommand{\da}{{\dot{\alpha}}}

\newcommand{\Db}{{\bar D}}

\newcommand{\cDb}{{\bar \cD}}

\newcommand{\Mb}{\bar{M}}

\newcommand{\vfb}{\bar{\varphi}}
\newcommand{\ct}{\tilde{c}}

\newcommand{\scG}{\mathscr{G}}

\newcommand{\scJ}{\mathscr{J}}

\newcommand{\rmd}{{\rm d}}
\newcommand{\rme}{{\rm e}}
\newcommand{\rmi}{{\rm i}}

\newcommand{\bm}[1]{\boldsymbol{#1}}
\newcommand{\non}{\nonumber}
\newcommand{\bem}{\begin{pmatrix}}
\newcommand{\eem}{\end{pmatrix}}
\newcommand{\eqn}[1]{(\ref{#1})} 
\def\dag{\dagger}
\newcommand{\pd}{\partial}
\newcommand{\ddt}[1]{\frac{\rmd #1}{\rmd t}}

\def\Tr{{\rm Tr}}
\def\tr{{\rm tr}}
\def\trF{{\rm tr}} 
\def\trad{{\rm tr}} 

\def\diag{{\rm diag}}
\def\cycl{{\rm cycl.}}
\def\sgn{{\rm sgn}}
\def\T{{\rm T}} 
\newcommand{\Li}{{\rm Li}}  
\newcommand{\su}[1]{\ensuremath{{\rm su}(#1)}}

\def\half{\ensuremath{\frac{1}{2}}}
\def\third{\ensuremath{\frac{1}{3}}}

\def\ord{\ensuremath{{\rm O}}}

\def\ie {{\rm ie~}}
\def\cc {{\rm c.c.}}

\def\ns {{\rm(no ~ sum)}}
\newcommand{\I}{{\rm I}}
\newcommand{\II}{{\rm II}}

\def\intk{\int\!\!\frac{\rmd^dk}{(2\p)^d}\,}
\def\intrho{\int\!\!{\rmd}^4\rho\,}
\def\intz{\int\!\!{\rmd}^8z\,}

\def\intc{\int\!\!{\rmd}^6z\,}
\def\intac{\int\!\!{\rmd}^6\bar z\,}

%% file: Gamma2.pdf_t
\begin{picture}(0,0)%
\includegraphics{Gamma2}%
\end{picture}%
\setlength{\unitlength}{4144sp}%
\begingroup\makeatletter\ifx\SetFigFontNFSS\undefined%
\gdef\SetFigFontNFSS#1#2#3#4#5{%
  \reset@font\fontsize{#1}{#2pt}%
  \fontfamily{#3}\fontseries{#4}\fontshape{#5}%
  \selectfont}%
\fi\endgroup%
\begin{picture}(2069,1487)(2229,-2154)
\put(3061,-1276){\makebox(0,0)[lb]{\smash{{\SetFigFontNFSS{12}{14.4}{\rmdefault}{\mddefault}{\updefault}{\color[rgb]{0,0,0}$\varphi_2$}%
}}}}
\put(3061,-1951){\makebox(0,0)[lb]{\smash{{\SetFigFontNFSS{12}{14.4}{\rmdefault}{\mddefault}{\updefault}{\color[rgb]{0,0,0}$\varphi_3$}%
}}}}
\put(3061,-961){\makebox(0,0)[lb]{\smash{{\SetFigFontNFSS{12}{14.4}{\rmdefault}{\mddefault}{\updefault}{\color[rgb]{0,0,0}$\varphi_1$}%
}}}}
\end{picture}%

%% file: Gamma1.pdf_t
\begin{picture}(0,0)%
\includegraphics{Gamma1}%
\end{picture}%
\setlength{\unitlength}{4144sp}%
\begingroup\makeatletter\ifx\SetFigFontNFSS\undefined%
\gdef\SetFigFontNFSS#1#2#3#4#5{%
  \reset@font\fontsize{#1}{#2pt}%
  \fontfamily{#3}\fontseries{#4}\fontshape{#5}%
  \selectfont}%
\fi\endgroup%
\begin{picture}(4769,1487)(2229,-2154)
\put(4501,-1523){\makebox(0,0)[lb]{\smash{{\SetFigFontNFSS{12}{14.4}{\rmdefault}{\mddefault}{\updefault}{\color[rgb]{0,0,0}$~+$}%
}}}}
\put(3106,-1996){\makebox(0,0)[lb]{\smash{{\SetFigFontNFSS{12}{14.4}{\rmdefault}{\mddefault}{\updefault}{\color[rgb]{0,0,0}$\varphi_{2}$}%
}}}}
\put(3106,-1006){\makebox(0,0)[lb]{\smash{{\SetFigFontNFSS{12}{14.4}{\rmdefault}{\mddefault}{\updefault}{\color[rgb]{0,0,0}$\varphi_{2}$}%
}}}}
\put(5806,-1006){\makebox(0,0)[lb]{\smash{{\SetFigFontNFSS{12}{14.4}{\rmdefault}{\mddefault}{\updefault}{\color[rgb]{0,0,0}$\varphi_{3}$}%
}}}}
\put(5806,-1996){\makebox(0,0)[lb]{\smash{{\SetFigFontNFSS{12}{14.4}{\rmdefault}{\mddefault}{\updefault}{\color[rgb]{0,0,0}$\varphi_{3}$}%
}}}}
\end{picture}%